# Equally Efficient Interlayer Exciton Relaxation and Improved Absorption in Epitaxial and Non-epitaxial MoS$_2$/WS$_2$ Heterostructures

Yifei Yu[1§], Shi Hu[1§], Liqin Su[3§], Lujun Huang[1], Yi Liu[4], Zhenghe Jin[5], Alexander A. Purezky[6], David B. Geohegan[6], Ki Wook Kim[5], Yong Zhang[3], Linyou Cao[1,2]*

[1]Department of Materials Science and Engineering, North Carolina State University, Raleigh NC 27695; [2]Department of Physics, North Carolina State University, Raleigh NC 27695; [3]Department of Electrical and Computer Engineering, The University of North Carolina at Charlotte, Charlotte, NC 28223; [4]Analytical Instrumentation Facility, North Carolina State University, Raleigh NC 27695; [5]Department of Electrical and Computer Engineering, North Carolina State University, Raleigh NC 27695; [6]Center for Nanophase Materials Sciences, Oak Ridge National Laboratory, Oak Ridge, Tennessee 37831

[§] These authors contribute equally.

**Abstract**

Semiconductor heterostructures provide a powerful platform to engineer the dynamics of excitons for fundamental and applied interests. However, the functionality of conventional semiconductor heterostructures is often limited by inefficient charge transfer across interfaces due to the interfacial imperfection caused by lattice mismatch. Here we demonstrate that MoS$_2$/WS$_2$ heterostructures consisting of monolayer MoS$_2$ and WS$_2$ stacked in the vertical direction can enable equally efficient interlayer exciton relaxation regardless the epitaxy and orientation of the stacking. This is manifested by a similar two orders of magnitude decrease of photoluminescence intensity in both epitaxial and non-epitaxial MoS$_2$/WS$_2$ heterostructures. Both heterostructures also show similarly improved absorption beyond the simple super-imposition of the absorptions of monolayer MoS$_2$ and WS$_2$. Our result indicates that 2D heterostructures bear significant implications for the development of photonic devices, in particular those requesting efficient exciton separation and strong light absorption, such as solar cells, photodetectors, modulators, and photocatalysts. It also suggests that the simple stacking of dissimilar 2D materials with random orientations is a viable strategy to fabricate complex functional 2D heterostructures, which would show similar optical functionality as the counterpart with perfect epitaxy.

Engineering the dynamics of excitons, including generation, dissociation, transfer, and recombination, by semiconductor heterostructures bears tremendous significance for fundamental and applied interests[1, 2]. It stands as a major strategy for the development of all kinds of devices that involve photon-to-electron or electron-to-photon conversions, such as solar cells, LEDs, lasers, photodetectors, modulators, and photocatalysts. It also provides platforms with well-controlled excitons for the studies of fundamental physics. However, the capability of conventional semiconductor heterostructures to engineer excitons is often limited by the difficulty in developing high-quality interfaces for efficient interfacial charge transfer, a key step for the engineering of excitons. Typical heterostructures consist of two or more dissimilar semiconductor materials, and a nice match between the crystalline lattices of the semiconductor materials is required to yield high-quality interfaces. This requirement of lattice match imposes a fundamental constraint for the design of the conventional semiconductor heterostructures with increasing compositional and structural complexity to provide sophisticated control of excitons.

Two-dimensional (2D) transition metal dichalcogenide (TMDC) materials in forms of a monolayer or fewlayer of atoms promise to enable a new type of semiconductor heterostructures. These materials present an atomic-scale semiconductor with bandgap in amplitude comparable to those of conventional group IV, III-V semiconductor materials. The heterostructures that consist of dissimilar 2D materials stacked in the vertical direction would provide capabilities to engineer excitons from a truly atomic level. Most importantly, unlike the conventional semiconductor heterostructures, which request lattice match to ensure high quality interfaces, 2D heterostructures may have high quality interfaces regardless substantial lattice mismatch[3-10]. This is because the interaction between 2D materials is van der Waals (vdW) forces and the weak interaction can relax the requirement of lattice match. Numerous works have recently demonstrated the fabrication of 2D heterostructures with the presence of lattice mismatch and the capability of the 2D heteostructures to efficiently engineer excito[11-20]. However, one very important question that has remained unanswered yet is how the excitonic properties of 2D heterostructures could depend on the epitaxy and orientation of the stacking. Knowledge of this question will provide useful guidance for the rational design of complex 2D heterostructures with desired exciton dynamics.

Here we have studied the excitonic properties of $MoS_2/WS_2$ heterostructures that consist of monolayer $MoS_2$ and $WS_2$ stacked either epitaxially or non-epitaxially in the vertical direction. Surprisingly, we demonstrate equally efficient interlayer relaxation of excitons in the heterostructures regardless the epitaxy and orientation of the stacking. This is manifested by a similar two-order magnitude decrease in the photoluminescence intensity of all the heterostructures compared to that of separate monolayers. Additionally, both epitaxial and non-epitaxial $MoS_2/WS_2$ heterostructures show similarly improved absorption that is beyond the simple super-imposition of the absorption of monolayer $MoS_2$ and $WS_2$, in particular for the incidence below the intrinsic bandgap of the monolayers. The non-epitaxial heterostructures are made by manually stacking single-crystalline monolayer $MoS_2$ and $WS_2$ pre-grown separately with a chemical vapor deposition (CVD) process reported previously[10, 21-25]. The epitaxial heterostructures, which are single crystalline as well, are synthesized by a CVD process that we have developed with a mixture of $MoO_3$ and $WO_3$ as the precursors (see Methods and Supporting Information). Our result indicates that 2D heterostructures bear significant implications for the development of photonic devices, particularly those requesting efficient exciton separation and strong light absorption, such as solar cells[5, 18-20], photodetectors, modulators, and photocatalysts. It also indicates that the simple stacking of dissimilar 2D materials with random stacking orientations may be a viable strategy to fabricate complex 2D heterostructures for the engineering of excitons.

We start the studies with epitaxial $MoS_2/WS_2$ heterostructures. Unlike non-epitaxial heterocturestures, whose band structure is difficult to theoretically predict due to the difficulty in building up unit cells in theoretical models, the band structure of epitaxial $MoS_2/WS_2$ heterostructures has been well studied using first principle techniques[11-14, 26]. This allows for synergistic studies from both experimental and theoretical sides to provide insights that are difficult to obtain from either perspective alone. The synthesized heterostructure consists of two concentric equilateral triangles in lateral size of tens of micrometers and well aligned in either the same or opposite directions (Figure 1a inset). Raman, STEM, and AFM characterizations indicate that the large triangle is single-crystalline monolayer $MoS_2$ while the small one single-crystalline monolayer $WS_2$, both continuous, smooth, and uniform (Figure 1a-b, and detailed characterizations seen in the Supporting Information). The STEM characterization also demonstrates that the $MoS_2$ and $WS_2$ monolayers, which have almost identical lattice constants[27, 28], are epitaxially stacked together in an A-B staking mode along the vertical direction (Figure1b and Figure S1-3).

We characterized the optical properties of the epitaxial $MoS_2/WS_2$ hetersotructure at room temperatures. Figure 2a shows the mapping of photoluminescence (PL) from a typical as-grown heterostructure whose optical image is given in Figure 2b. The structure consists of a small $WS_2$ monolayer in lateral size of ~ 8μm epitaxially stacked on the top of a big $MoS_2$ monolayer in lateral size of ~ 25μm. We can immediately find that the PL from the edge region, which corresponds to monolayer $MoS_2$, is much stronger than that from the center where the $MoS_2/WS_2$ heterostructure is located. Representative PL spectra extracted from the mapping results are plotted in Figure 2c. The PL spectrum collected from the monolayer $MoS_2$ region (the big triangle) exhibits a strong peak at 1.87 eV, consistent with what was found for monolayer $MoS_2$ previously[29]. The PL collected from the $MoS_2/WS_2$ heterostructure region ( the small triangle) shows a peak position similar to that of the $MoS_2$, but its intensity is smaller than that of the monolayer $MoS_2$ by two orders of magnitude (Figure 2c). To further illustrate the low PL efficiency of the heterostructure, the PL of bilayer $MoS_2$ collected under comparable conditions is given in Figure 2c as well (also see Figure S4). It has been well known that bilayer $MoS_2$ shows weaker PL than monolayer $MoS_2$ because the transition of the bandgap from direct to indirect[29-32]. However, the PL intensity of the $MoS_2/WS_2$ heterostructure can be found even much weaker than that of bilayer $MoS_2$. We would like to point out two differences of our results from those of one recent study[18] on similar epitaxial $MoS_2/WS_2$ heterostructures that was published during the review process of this work. First, the PL intensity of epitaxial $MoS_2/WS_2$ heterostructures in that study shows only three times smaller than that of monolayer $MoS_2$, instead of a two orders of magnitude decrease as we observed. Second, in that study an additional PL peak at 1.4 eV was reported resulting from interlayer exciton transition, but we did not observe this PL peak in our materials even at a low temperature of 10 K (Figure S4). There are two possible reasons accounting for the differences. One could be the different synthetic processes, which might cause some differences in the resulting materials. Unlike that study, which used a mixture of element tungsten and tellurium as the precursor for $WS_2$, we used $WO_3$ instead. The other reason could be the difference in substrates, as sapphire substrates were used in our experiments while $SiO_2/Si$ substrates used in that study. It has been well known that substrates could substantially affect the PL of 2D materials[33].

The established theoretical calculations for the bandstructure of epitxial MoS$_2$/WS$_2$ heterostructures can provide useful insights into the fundamental physics underlying the observed low PL efficiency[11-14, 26]. Theoretical calculations have indicated that the band structure of epitaxial MoS$_2$/WS$_2$ heterostructures at the *K* point in the Brillouin zone is approximately a simple superposition of the states of monolayer MoS$_2$ and WS$_2$[11, 12, 26]. The MoS$_2$/WS$_2$ heterostructure essentially makes a type II heterojunction with the valance band maximum (VBM) completely localized to WS$_2$ and the conduction band minimum (CBM) to MoS$_2$[11, 12, 26]. Indeed, we observed similar PL peak positions in the heterostructure and monolayer MoS$_2$ in experiments (Figure 2c), and this supports the theoretical prediction that the bandstructures at the *K* point may not change much after the heterostructuring. As a result, we conclude that the observed low PL of the heterostructure is due to the interlayer relaxation (dissociation) of excitons as illustrated in Figure 2d. The band structure offset between the MoS$_2$ and WS$_2$ monolayers can facilitate the separation of photo-excited charges, electrons to MoS$_2$ while holes to WS$_2$. This separation in different monolayers decreases the spatial overlap between the wave functions of electrons and holes, which may subsequently lead to the decrease in PL efficiency[34]. The observed low PL efficiency also indicates that the interlayer relaxation (dissociation) is very fast. As illustrated in Figure 2d, the interlayer relaxation of the photo-excited charges at band edges competes with another relaxation pathway, intralayer recombination. Given the simple superposition of the band structure as theoretically predicted[11, 12, 24], it is reasonable to assume that the intralayer recombination (at the *K* point) of the heterostructure is similar to that of standing-alone monolayers. The result that the PL of the heterostructure is 50-100 times weaker than that of monolayers MoS$_2$ implies that the interlayer relaxation process is 50-100 times faster than the intralayer recombination in monolayer MoS$_2$. We can further roughly estimate the interlayer relaxation to be in a timescale of 10-100 fs as the intralayer recombination of excitons in monolayer $MoS_2$ is reported in scale of around 1-5 ps[35, 36]. This estimate is reasonably consistent with the result of another recent study[17] that was published during the review process of this work, in which the interlayer transfer process in $MoS_2/WS_2$ heterostructures is experimentally measured to be within 50 fs. Note that the dramatic decrease in the PL efficiency of epitaxial $MoS_2/WS_2$ heterostructures we observed in experiments is actually different from what predicted in theory. The theoretical calculation did not fully recognize how efficient the interlayer relaxation could be and predicted substantial PL signal in the heterostruecture due to the presence of direct transition at the K point[11, 12].

Very surprisingly, to achieve the efficient interlayer relaxation of excitons does not require the heterostructure to be epitaxially stacked. We observed similarly efficient interlayer relaxation of excitons in non-epitaxial $MoS_2/WS_2$ heterostructures as well. To make the non-epitaxial $MoS_2/WS_2$ heterostructure, we first grew single-crystalline $MoS_2$ and $WS_2$ monolayers separately on sapphire substrates using the CVD processes reported previously[23, 37] and then transferred the monolayer $MoS_2$ onto the top of monolayer $WS_2$ using a unique surface energy-assisted transfer approach that we have recently developed[38] (see Methods and Figure S5). Different from the transfer techniques used in the previous works for the fabrication of heterostructures[17, 23, 26], which involved chemical etchants and would most likely cause damages and leave organic residues at the transferred materials, our surface energy-assisted approach relies on room temperature water droplet to transfer the monolayers and are proved able to better protect the quality of the transferred materials with no damage and organic residues left behind[38]. After the transfer, we mildly annealed the heterostrucrure at 200-250 °C for 10-30 minutes under an Ar flow to remove solvent or water residues. We have confirmed that both $MoS_2$ and $WS_2$ monolayers are very stable and this mild annealing process cannot cause any change in the quality and crystalline structures of the materials.

Figure 3a-b shows the result of PL mapping for a typical non-epitaxial $MoS_2/WS_2$ heterostructure and the optical image of the heterostructures mapped. The heterostructures consist of numerous small single-crystalline monolayer $MoS_2$ in lateral size of ~ 5μm randomly stacked on top of a big single-crystalline $WS_2$ monolayer in size of ~ 50μm (also see Figure S6). Similar to what we find with the epitaxial $MoS_2/WS_2$ heterostructure, the PL of all the non-epitaxial $MoS_2/WS_2$ heterostructures is two orders of magnitude weaker than that of monolayer $MoS_2$ or $WS_2$, regardless the relative orientation of the monolayers (Figure 3b-c), indicating the general presence of efficiency interlayer exciton relaxation in all the heterostructures. The interlayer relaxation process is very sensitive to the surface quality of the heterostructures that may affect the coupling between the $MoS_2$ and $WS_2$ monolayers. We find that the non-epitaxial $MoS_2/WS_2$ heterostructures without being treated by the low temperature annealing (gray curve in Figure 3c) show a less decrease in PL intensity than the annealed one. The low-temperature annealing process may remove the residue of solvent and water molecules left between the two monolayers during the transfer process, which may subsequently facilitate the interlayer exciton relaxation. The independence of the efficient interlayer exciton relaxation in $MoS_2/WS_2$ heterostructures on the epitaxy and orientation of the stacking suggests a strong electron-phonon coupling in 2D materials[39]. The electron-photon coupling could be so strong that able to efficiently compensate whatever momentum mismatch for the charge transfer between the monolayers. A full-fledged study on the electron-phonon coupling is beyond the scope of this work.

To further understand our conclusion, we compare our results with what have been recently published during the review of this work[17, 18, 26]. The two orders of magnitude decrease in PL intensity we observed is substantially greater than what reported by the other groups for non-epitaxial $MoS_2/WS_2$ heterostructures [17, 18, 26], which typically see a decrease less than 3 times. The difference could be due to the different transferring processes used in the fabrication of the heterostructure. The unique surface energy-assisted transfer we used[38] can better protect the quality of the transferred monolayer with no organic residue and damage left than the approach used by the other groups[17, 18, 26]. This could result in better coupling between the two monolayers involved and hence higher efficiency of interlayer exciton relaxation in the non-epitaxial heterostructure we made. Another possible reason could be related with the substrates. The two-order magnitude decreases in PL intensities we observed is from the non-epitaxial $MoS_2/WS_2$ hetersotructures on sapphire substrates, while the non-epitaxial $MoS_2/WS_2$ hetersotructures studied in some of the previous works[18, 26] were fabricated on $SiO_2/Si$ substrates. We did also observe a lesser decrease in PL at the non-epitaxial $MoS_2/WS_2$ heterostructures made on $SiO_2/Si$ substrates (Figure S7). Nevertheless, the different decrease in the PL caused by using different substrates does not affect the generality of our conclusion. For instance, the decrease in PL that we observed at the non-epitaxial $MoS_2/WS_2$ heterostructures made on $SiO_2/Si$ substrates (Figure S7) is very similar to the PL decrease recently observed at the epitaxial $MoS_2/WS_2$ heterostructures grown on $SiO_2/Si$ substrates[18]. This suggests that our conclusion for the independence of the interlayer exciton relaxation on the stacking epitaxy and orientation can be generally applied to the $MoS_2/WS_2$ heterostructures on other substrates, although using different substrates might lead to different absolute amplitudes of the decrease in PL

Whereas the PL is dramatically suppressed due to efficient interlayer exciton relaxation, $MoS_2/WS_2$ heterostructures show improved absorption that is beyond a simple superimposition of the absorptions of monolayer $MoS_2$ and $WS_2$. The absorption improvement is particularly prominent for the incidence below the intrinsic band gap of the monolayers. The absorption spectra measured from epitaxial, annealed non-epitaxial, and non-annealed non-epitaxial $MoS_2/WS_2$ heterostructures are plotted in the upper panels of Figure 4 along with the absorption spectra of corresponding monolayer $MoS_2$. We can find that the epitaxial and annealed non-epitaxial heterostructures exhibit substantially higher absorption efficiencies than monolayer $MoS_2$ for the incidence below the bandgap of $MoS_2$, which is to the left of the dashed red lines (Figure 4a-b upper panels). But similar absorption improvement for the sub-bandgap incidence cannot be found in the non-epitaxial heterostructure without being annealed (Figure 4c upper panel). To further illustrate the relationship between the absorptions of the heterostructure and the monolayers involved, we subtract the absorption spectra of monolayer $MoS_2$ from those of $MoS_2/WS_2$ heterostructures. For the non-annealed non-epitaxial heterostructures, the spectrum resulted from the subtraction is identical to the absorption spectrum of monolayer $WS_2$ (Figure 4c lower panel), indicating the absorption of the heterostructure is a simple superposition of separate monolayer $MoS_2$ and $WS_2$. But for the epitaxial and annealed non-epitaxial heterostructures, the subtraction gives rise to a peak at 1.84 eV along with the features resulting from the absorption of monolayer $WS_2$ (Figure 4a-b lower panels). Intuitively, this extra peak, which indicates improved absorption in the heterostructure for the low energy incidence, results from the red-shift of exciton peaks due to the reduction of dielectric screening in the heterostructure, similar to what observed in bilayer $MoS_2$ and bilayer $WS_2$ (Figure S8) [40].

Our result indicates that 2D heterostructures present a useful platform for the engineering of excitons at the atomic level. For instance, it provides the capabilities to efficiently dissociate the excitons in 2D materials that would otherwise be difficult to separate and tend to radiatively recombine due to extraordinarily strong excitonic binding energy[41, 42]. The combination of efficient exciton dissociation and improved absorption make 2D heterostructures particularly useful for the absorption-based photonic devices, such as photovoltaics, solar fuels, photodetectors, optical modulators, and photocatalysts. Additionally, the independence of the interlayer exciton relaxation on the stacking epitaxy and orientation clearly points out that the simple stacking of 2D materials in the vertical direction with random orientations is a viable strategy for the fabrication of functional 2D heterostrctures. Complex 2D heterostructures fabricated by manually stacking dissimilar 2D materials with random orientation may show equal optical functionality as the counterpart with perfect epitaxy.

**Methods**

*Synthesis of epitaxial $MoS_2/WS_2$ heterostructures.* The epitaxial heterostructures were synthesized by using a chemical vapor deposition process that we have developed by adapting what was previously reported for the synthesis of monolayer $MoS_2$ and $WS_2$[23, 37]. Briefly, the synthesis was performed in a tube furnace with sulfur (typically 1.0g) and a mixture of $MoO_3$ and $WO_3$ (typically 80 mg with the weight percent of $MoO_3$ 1% and $WO_3$ 99%) as the precursors. The sulfur was placed at the upstream of the tube furnace and the mixed $MoO_3/WO_3$ at the center. Sapphire substrates were placed at the downstream in the tube. Other typical experimental conditions including a temperature of 950 °C and a flow of Ar gas in a rate of ~ 100 sccm. In a typical synthetic process, the temperature was ramped to 950 °C in 35 min and kept 950 °C for 2 hours. After that, the whole system was cooled down to room temperature naturally.

*Fabrication of non-epitaxial MoS$_2$/WS$_2$ heterostructures.* The non-expitaxial heterostructures were made by manually stacking monolayer MoS$_2$ and WS$_2$ that were grown using the chemical vapor deposition process reported previously[23, 37]. The process is similar to what was used for the synthesis of heterostructures, but only either MoO$_3$ or WO$_3$ instead of the mixture was used for the synthesis of MoS$_2$ or WS$_2$. The typical temperatures used for the synthesis of MoS$_2$ is 750°C and the temperature for WS$_2$ 900°C. Ar was used as the carrier gas for the synthesis of MoS$_2$ and forming gas (5% H$_2$ in Ar) for WS$_2$ with a flux rate of 100 sccm in both cases.

To make the non-epitaxial heterostructure, we first lifted off the synthesized MoS$_2$ from the growth substrate using a surface energy-assisted transfer process that we have recently developed[38]. Briefly, in a typical transfer process, 9g of polystyrene (PS) with a molecular weight of 280,000 g/mol was dissolved in 100 ml toluene and then the PS solution was spin coated (3500 rpm for 60s) on the as-grown MoS$_2$ on sapphire substrates. This was followed by a baking at 80-90 °C for 15 minutes). A water droplet was then dropped on top of the sample. To facilitate the penetration of water molecules, we poked the PS layer with a sharp object from the edge. Once the PS layer was scratched from the edge, water molecules could penetrate through all the way under MoS$_2$, resulting the delamination of the PS-MoS$_2$ assembly. The water droplet was then removed away with paper towel. We could pick up the polymer/ MoS$_2$ assembly with a tweezers and transferred it onto as-grown WS$_2$ monolayers. To ensure the uniformity of the transferred MoS$_2$, we baked the transferred PS-MoS$_2$ assembly at 80 °C for 1 hour and a final baking for 30 mins at 150 °C. Finally, PS was removed by rinsing with toluene several times.

After that, the stacked heterostructure was annealed at 200-250 °C for 10-30 minutes under Ar environment.

*Characterizations of $MoS_2$ films.* High resolution STEM images were taken using the FEI Titan 80-300 probe aberration corrected and monochromated Scanning Transmission Electron Microscope (STEM) operated at 200 kV. In STEM mode, Z-contrast images were taken using a high-angle annular dark-field (HAADF) detector (Fischione Instrument), and elemental mapping was performed using the "Super X" Energy Dispersive Spectrometric (EDS) system. We transferred the synthesized materials to TEM grids following a surface energy-assisted transfer approach that we have recently developed[38]. The convergence angle was set at 21 mrad, and probe current was about 110 pA, at which, we found that the beam damage on the $MoS_2$ sample can be controlled to the minimum within a reasonable period of time for imaging. The thickness and surface topology were measured using atomic force microscope (AFM, Veeco Dimension-3000). Raman and photoluminescence (PL) measurements were carried out using Horiba Labram HR800 system using a 532 nm laser.

A home-built setup that consists of a confocal microscope (Nikon Eclipse C1) connected with a monochromator (SpectraPro, Princeton Instruments) and a detector (Pixis, Princeton Instruments) was used to perform the absorption measurement. In a typical measurement, we collected the white light transmitted through the sample using a 100× objective with a numerical aperture of 0.9 (Nikon). The light was from a halogen lamp and was broadly cast onto the samples with no focusing. We obtained the spectral absorption efficiency by normalized the transmitted light with $I_1$ and without the sample $I_0$ as $(I_0-I_1)/I_0$. A focal plane aperture at the confocal scanning head installed with the microscope allows us to define the sample area to be measured with a spatial resolution of 300 nm.

**Figure Captions**

**Figure 1. Characterizations of epitaxial $MoS_2/WS_2$ heterostructures.** (a) Raman spectra collected from different areas (region 1 and 2) of the epitaxial heterostructure. The assignment for the Raman peaks is given as shown. Inset, optical images of two typical epitaxial $MoS_2/WS_2$ heterostructures with different relative orientations. The larger triangle is monolayer $MoS_2$ while the small one at the center is monolayer $WS_2$. Scale bar, 10 μm. (b) Scanning transmission electron microscope high angle annular dark field image (STEM-HAADF) of the epitaxial heterostructure. The W and Mo atoms, which show different contrasts, are denoted in the figure. The lattice constant is measured in the figure as well. The dashed orange lines indicate the crystalline directions of the $WS_2$ layer. The circles in orange and blue represent W and Mo atoms, respectively, which are used to illustrate the offset of these atoms. Inset, the Fast Fourier transformation pattern of the image.

**Figure 2. Low PL efficiency of epitaxial $MoS_2/WS_2$ heterostructures.** (a) PL mapping of a typical epitaxial $MoS_2/WS_2$ heterostructure. (b) Optical image of the heterostructure mapped in (a). (c) Spectra PL collected from the monolayer (1L) $MoS_2$ area (red curve) and the $MoS_2/WS_2$ area (blue curve) of the heterostructure. The PL from a $MoS_2$ bilayer (2L) is also given (black curve). Inset, comparison of the PL from the $MoS_2$ area and the $MoS_2/WS_2$ area, where the PL from the $MoS_2/WS_2$ area scaled by a factor of 60 for visual convenience. (d) Schematic

illustration for the bandstructure alignment of the heterostructure. The *K* point of MoS$_2$ coincides with the *K'* point of WS$_2$. The interlayer relaxations and intralayer recombination are illustrated.

**Figure 3. Low PL efficiency of non-epitaxial MoS$_2$/WS$_2$ heterostructures.** (a) PL mapping of typical non-epitaxial MoS$_2$/WS$_2$ heterostructures. (b) Optical image of the heterostructure mapped in (a). It consists of mutliple monolayer MoS$_2$ (small triangles) randomly distributed on top of a big monolayer WS$_2$. This can be seen more clearly in the image given in Figure S6 that shows the edge of the monolayer WS$_2$. (c) Spectra PL collected from the non-epitaxial MoS$_2$/WS$_2$ heterostructure (red curve), monolayer (1L) MoS$_2$ (blue curve), and monolayer (1L) WS$_2$ (brown curve). The PL from the non-epitaxial MoS$_2$/WS$_2$ heterostructure without being annealed is also given (grey curve).

**Figure 4. Improved absorption of MoS$_2$/WS$_2$ heterostructures.** Absorption spectra collected from the MoS$_2$ area (blue curve) and the MoS$_2$/WS$_2$ area (black curve) for (a) epitaxial MoS$_2$/WS$_2$ heterostructures, (b) annealed non-epitaxial MoS$_2$/WS$_2$ heterostructures, and (c) non-epitaxial MoS$_2$/WS$_2$ heterostructure without being annealed. The dashed red lines indicate the boundary where the epitaxial and annealed non-epitaxial heterostructures show obvious higher absorption for the incidence to the left than the monolayer MoS$_2$. The difference between the two absorption spectra (MoS$_2$/WS$_2$ - MoS$_2$) is given in the corresponding lower panel (red curve). For the non-epitaxial heterostructures, the absorption spectrum of monolayer WS$_2$ (brown curve)

is also given in the lower panel as a reference. The black arrows point towards the peaks indicating the improved absorption of the heterostructures.

## ASSOCIATED CONTENT

**Supporting Information**

(1) Detailed experimental results on the synthesis and compositional and structural characterizations of the synthesized epitaxial $MoS_2/WS_2$ heterostructures; (2) more Raman and optical measurements for bilayer $MoS_2$, bilayer $WS_2$, and $MoS_2/WS_2$ heterostructures. This material is available free of charge *via* the Internet at http://pubs.acs.org.

## AUTHOR INFORMATION

**Corresponding Authors**

*lcao2@ncsu.edu

**Author Contributions**

§These authors contribute equally.

**Notes**

The authors declare no competing financial interests.

# ACKNOWLEDGEMENTS

This work was supported by a Young Investigator Award from the Army Research Office (W911NF-13-1-0201) and, partially, by a CAREER award from the National Science Foundation (DMR- 1352028). K. W. K. acknowledges the support from FAME (one of six centers of STARnet, a SRC program sponsored by MARCO and DARPA). Y. Z. acknowledges the support of the Bissell Distinguished Professorship. The authors acknowledge the use of the Analytical Instrumentation Facility (AIF) at North Carolina State University, which is supported by the State of North Carolina and the National Science Foundation. Part of the Raman and PL work was conducted at the Center for Nanophase Materials Sciences, which is sponsored at Oak Ridge National Laboratory by the Scientific User Facilities Division, Office of Basic Energy Sciences, U.S. Department of Energy..

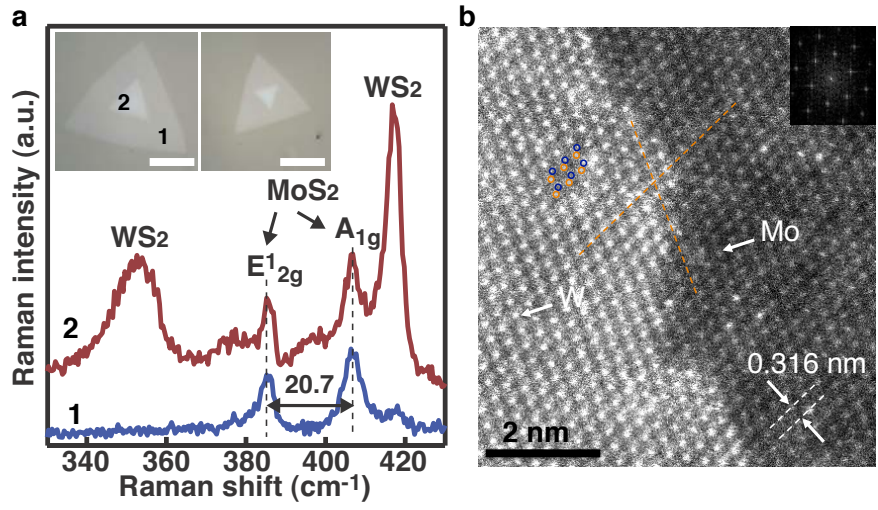

**Figure 1. Yu et al.**

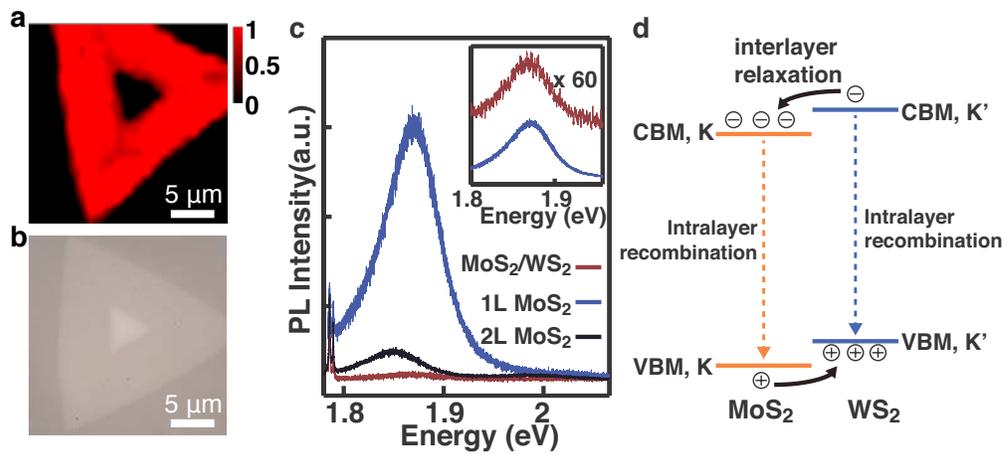

**Figure 2. Yu et al.**

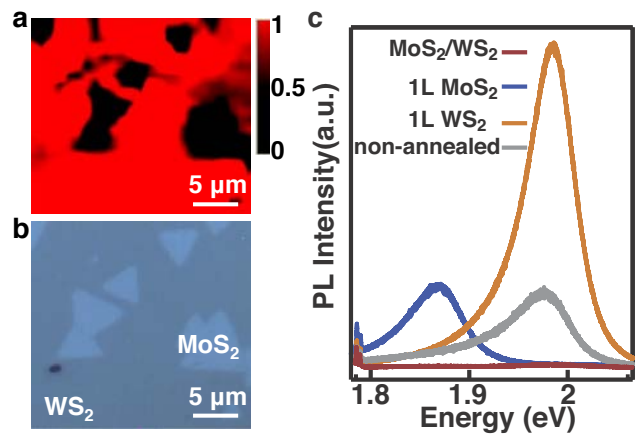

**Figure 3. Yu et al.**

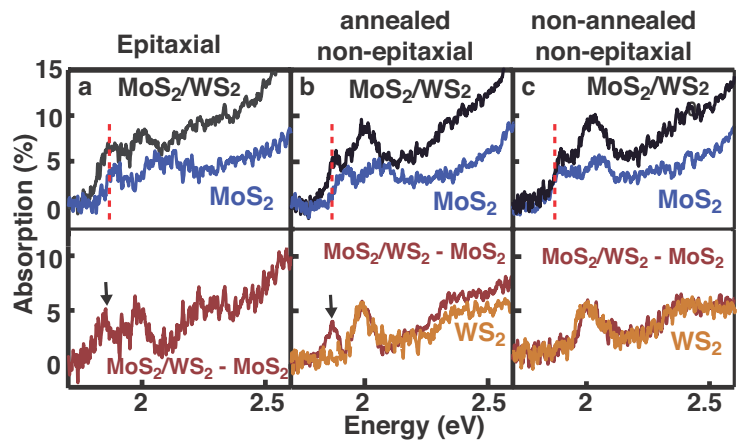

**Figure 4. Yu et al.**

**TOC**

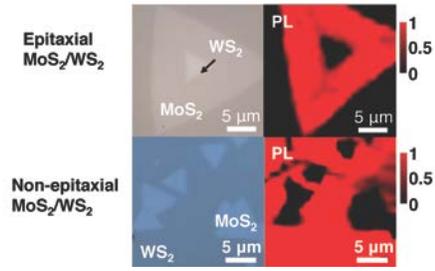